\documentclass[aps,prl,twocolumn,superscriptaddress,floatfix]{revtex4-2}
\usepackage{amsmath,amssymb,bm,graphicx,mathtools,csquotes}
\usepackage[colorlinks=true,allcolors=blue]{hyperref}
\usepackage{microtype}
\usepackage{float}

\begin{document}
\title{Onset of subdiffusion in dephased quasiperiodic lattices}

\author{Ashlin V Thomas}
\thanks{These authors contributed equally.}
\affiliation{School of Physical Sciences, National Institute of Science Education and Research, Jatni 752050, India}
\affiliation{Homi Bhabha National Institute, Training School Complex, Anushaktinagar, Mumbai 400094, India}

\author{Sanchayan Banerjee}
\thanks{These authors contributed equally.}
\affiliation{School of Physical Sciences, National Institute of Science Education and Research, Jatni 752050, India}
\affiliation{Homi Bhabha National Institute, Training School Complex, Anushaktinagar, Mumbai 400094, India}

\author{Tapan Mishra}
\email{mishratapan@niser.ac.in}
\affiliation{School of Physical Sciences, National Institute of Science Education and Research, Jatni 752050, India}
\affiliation{Homi Bhabha National Institute, Training School Complex, Anushaktinagar, Mumbai 400094, India}

\begin{abstract}
Dephasing is widely expected to induce diffusion in quasiperiodic lattices due to the loss of phase coherence. However, this expectation can fail dramatically in lattices possessing singular onsite potentials or vanishing hoppings which are expected to trivially localize the particle. Remarkably, we show that the transport in such lattices can be subdiffusive under proper conditions for which the \enquote{mean resistance} of the lattice diverges. As a result, we obtain a transition from diffusive to subdiffusive transport depending on the strength of quasiperiodic modulation.  
Even more strikingly, the transport is slowest not deep in the subdiffusive regime but precisely at its onset, forming a sharp \enquote{subdiffusive trench}. Our analytical and numerical analysis reveal that the transport exponents are universal, fixed solely by the order of singularities (zeros) of the diagonal (off-diagonal) quasiperiodic modulation. In the end, we complement our findings through exact Lindblad simulations and provide possible experimental routes to realize such non-trivial transport. 
\end{abstract}

\maketitle

\textit{Introduction}---The interplay between disorder and quantum interference gives rise to localization of quantum states and suppression of transport - a hallmark feature of disordered quantum lattices~\cite{Anderson1958,EversMirlin2008}.
Central to this phenomenon is the phase coherence  of the particle's wave function across lattice sites, which in realistic settings is degraded when the system is coupled to an environment. This dephasing noise scrambles relative phases between lattice sites which destroys the phase coherence necessary for the quantum interference and hence unfreezes localized states \cite{moix2013coherent,Logan1987Dephasing} resulting in the restoration of ergodicity and  transport~\cite{Rebentrost2008EnvironmentAssisted,HakenStrobl1973,Plenio2008Dephasing,Caruso2009Highly}. Such an environment or dephasing assisted quantum transport has been a topic of paramount interest in recent years due to its role in enhancing performance of thermal machines~\cite{Lacerda2021,Chiaracane2022}, excitation transfer in photosynthetic complexes~\cite{Mohseni2008,Plenio2008Dephasing,Rebentrost2008EnvironmentAssisted,Caruso2009Highly} and bio-inspired solar-energy conversion~\cite{Romero2017}.

Understanding the nature of the transport due to competing disorder and dephasing effects has attracted a great deal of attention in recent years. Efforts have been made in this direction by considering different disorder lattices subject to dephasing noise. While the transport in presence of dephasing is often shown to be diffusive in nature~\cite{Znidaric2012Transport,Amir2009Classical,Longhi2024}, instances of subdiffusive transport have been reported under more restrictive conditions: when dephasing acts on a subset of sites~\cite{TaylorScardicchio2021}; at intermediate times, for slowly fluctuating noise~\cite{Geraldi2021Transient,Gopalakrishnan2017} or in a dephased dimerized off-diagonal quasiperiodic chain~\cite{Tao2026}; or in the motion of the wave packet's center of mass, while the ensemble-averaged variance still grows diffusively~\cite{Longhi2024Subdiffusive}. Although these studies have revealed that dephasing and localization oppose each other, the general criterion for the cessation of diffusion in a disordered lattice under uniform dephasing, its microscopic mechanism, and the nature of transport in the absence of diffusion have remained unclear.

\begin{figure}[t]
\centering
\includegraphics[width=\linewidth]{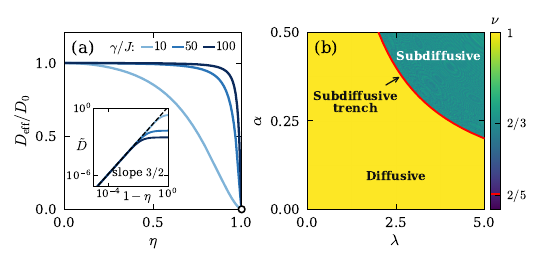}
\caption{(a) Effective diffusion coefficient $D_{\mathrm{eff}}/D_0$ [Eq.~\eqref{eq:Deff_analytical}] for onsite modulation versus $\eta = \lambda\alpha$ for dephasing rate $\gamma=10J, ~ 50J$ and $100J$ (light to deep blue), with $\alpha=0.2$ and $D_0=2J^2/\gamma$. The open circle at $(1,0)$ marks the one-sided limit $D_{\mathrm{eff}}\to0$ as $\lambda\alpha\to1^-$. Inset: $\widetilde D=(D_{\mathrm{eff}}/D_0)/(\gamma\alpha)^2$ collapses onto $\sqrt2\,(1-\lambda\alpha)^{3/2}$ (dashed line) as $\eta\to1^-$. (b) Transport phase diagram in the $(\lambda,\alpha)$ plane for $\gamma=10J$, displaying $\nu=2/z$ obtained from the rate-matrix gap $\Delta_W\propto L^{-z}$ for system sizes $L=610,987,1597,2584,4181$, averaged over 50 values of $\phi$. The diffusive ($\lambda\alpha<1$, $\nu=1$) and subdiffusive ($\lambda\alpha>1$, $\nu\approx2/3$) phases are separated by a sharp subdiffusive trench along the hyperbola $\eta=1$ (red line), where $\nu\approx2/5$.}
\label{fig:1}
\end{figure}

In this Letter, we answer this fundamental question for one-dimensional chains with either diagonal or off diagonal quasiperiodic disorder subjected to onsite dephasing. Although diffusion is the generic expectation under strong dephasing, we show that it survives if and only if the mean resistance, that is, the average inverse transfer rate, remains finite in the thermodynamic limit. Diffusion breaks down when the mean resistance becomes infinite, which happens at singularities (zeros)  of the onsite potential (hopping). Below the threshold onsite (hopping) modulation strength at which such a singularity (zero) first appears, the transport is diffusive, with a diffusion coefficient that vanishes continuously as the threshold is approached [Fig~\ref{fig:1}(a)]. At and beyond the threshold, vanishing transfer rates would naively be expected to cut the chain and trivially localize the particle~\cite{Longhi2024}. Remarkably, we show that the transport instead remains subdiffusive as the transfer rates become arbitrarily small but generically never vanish.
 
Surprisingly, exactly at the threshold, the transport is the slowest. Accordingly transport exponent $\nu$, defined by $\langle \Delta x^2(t)\rangle\sim t^{\nu}$, responds non-monotonically: it plunges from its diffusive value $\nu=1$ to a minimum $\nu=2/5$ at the threshold, forming a critical ``subdiffusive trench,'' and then recovers to $\nu=2/3$ beyond it, where transport remains subdiffusive [Fig~\ref{fig:1}(b)]. 
We show that these exponents are universal and are fixed solely by the order of the zeros of the transfer rates, independent of the other details of the modulation and of the dephasing rate. To rule out artifacts of the quantum-to-classical mapping, we further simulate the full Lindblad dynamics, which displays long-lived anomalous spreading consistent with these predictions.
 
\textit{Model}---We consider the one-dimensional Hamiltonian
\begin{equation}
\hat{H}=-\sum_j J_j\bigl(\hat c_j^\dagger\hat c_{j+1}+\mathrm{H.c.}\bigr)+\sum_jV_j\hat c_j^\dagger\hat c_j ,
\label{eq:Hamiltonian}
\end{equation}
where $\hat c_j^\dagger$ ($\hat c_j$) creates (annihilates) a particle at site $j$, $J_j$ is the hopping strength on bond $(j,j+1)$ and $V$ is the onsite potential at site $j$. In order to study the effect of dephasing we consider two different quasiperiodic lattice modulations; one by modulating the onsite potential and the other by modulating the hopping. In the first case, we assume the hopping to be uniform, $J_j=J$, and the onsite potential modulation
\begin{equation}
V_j=\frac{\lambda\cos(2\pi\beta j+\phi)}{1-\lambda\alpha\cos(2\pi\beta j+\phi)},
\label{eq: potential1}
\end{equation}
with $\lambda,\alpha>0$ and $\beta=(\sqrt5-1)/2$, the inverse golden ratio, and a quasiperiodic phase $\phi\in[0,2\pi)$. This model was first proposed by  some of us in Ref.~\cite{Banerjee2025} which is a variant of the generalized Aubry-Andr\'e potential of Ganeshan, Pixley, and Das Sarma~\cite{Ganeshan2015,Grempel1982,Avila2015,Biddle2011}, where $\alpha$ in the denominator is replaced by $\lambda\alpha$ which we denote as $\eta$. 
The interesting feature of this potential is that it is bounded for $\eta<1$, develops a double pole at $\eta=1$, and has two simple poles, at $2\pi\beta j+\phi=\pm\arccos(1/\eta)$, for $\eta>1$. As the second case, we consider the off-diagonal Aubry-Andr\'e model~\cite{Liu2018PhaseDiagram,Kraus2012Topological,Cestari2016Fate,vatnik2017anderson}, with $V_j=0$ and $J_j=A+B\cos(2\pi\beta j+\phi)$, where $A,B>0$ and $\kappa=B/A$. In close analogy, the hopping is bounded away from zero for $\kappa<1$, develops a quadratic zero at $\kappa=1$, and has two simple zeros, at $2\pi\beta j+\phi=\pi\pm\arccos(1/\kappa)$, for $\kappa>1$, where the closed system hosts critical, multifractal eigenstates~\cite{Xiao2021,Meier2016Atom}.
 
With this setup in hand, we proceed in three steps. We first couple the chain to a uniform dephasing environment and, in the strong-dephasing limit, map the quantum dynamics onto a classical random walk. Viewing this walk as a chain of resistors in series, we derive a general criterion for diffusion and the exponents that govern transport when it fails. Finally, we apply them to the two models and confirm these predictions with simulations of the full Lindblad dynamics.

The local dephasing is described by the Lindblad master equation~\cite{Lindblad1976,Gorini1976,Breuer2002}
\begin{equation}
\frac{d\hat\rho}{dt}=-i[\hat H,\hat\rho]+\gamma\sum_j\left(\hat n_j\hat\rho\hat n_j-\frac12\{\hat n_j,\hat\rho\}\right),
\label{eq:LME}
\end{equation}
with $\hat n_j=\hat c_j^\dagger\hat c_j$, dephasing rate $\gamma$, and $\hbar=1$. Since particle number is conserved in the system, we work in the single-particle sector. For $\gamma \gg J_j$, elimination of the rapidly decaying coherences~\cite{HakenStrobl1973,Znidaric2012Transport,Ray2026,Longhi2024} yields a classical master equation for the site populations $p_j=\rho_{jj}$,
\begin{equation}
\dot p_j=W_{j,j+1}(p_{j+1}-p_j)+W_{j,j-1}(p_{j-1}-p_j),
\label{eq:classical_master_eq}
\end{equation}
with the transition rate:
\begin{equation}
W_{j,j+1}=W_{j+1,j}=\frac{2J_j^2\gamma}{\gamma^2+(V_j-V_{j+1})^2}.
\label{eq:hopping_rates}
\end{equation}

Equation~\eqref{eq:classical_master_eq} describes a classical continuous-time random walk on a one-dimensional lattice with site-dependent hopping rates~\cite{Alexander1981,zwanzig1982non,Bouchaud1990Anomalous}.  In matrix form, $d\mathbf{p}/dt=W\mathbf{p}$, where the rate matrix $W$ is real symmetric and negative semidefinite. The spectral gap $\Delta_W$ (the magnitude of the lowest nonzero eigenvalue of $W$) governs the asymptotic relaxation time $\tau_{\rm rel}\sim1/\Delta_W$.

Equation~\eqref{eq:hopping_rates} shows that zeros of the transition rate can arise in only two ways: the denominator diverges, through singularities of the onsite potential, or the numerator vanishes, through zeros of the hopping. We first derive the transport properties for a generic model in Eq.~\ref{eq:Hamiltonian} using the resistance criterion discussed below, and then apply them to the two specific quasiperiodic models.
 
\textit{Resistance criterion and exponents}---In one dimension, the inverse transition rates add in series~\cite{Bernasconi1978,HausKehr1987,zwanzig1982non,Derrida1983,Alexander1981}, so the diffusion coefficient is the inverse mean bond resistance,
\begin{equation}
D_{\mathrm{eff}}=\left( \lim_{L\to\infty}\frac1{L-1}\sum_{j=1}^{L-1}\frac1{W_{j,j+1}} \right)^{-1} \equiv\Bigl\langle\frac1W\Bigr\rangle^{-1}.
\label{eq:Deff_general}
\end{equation}
Diffusion therefore survives if and only if $\langle1/W\rangle$ remains finite in the thermodynamic limit.
On a quasiperiodic lattice, the transfer rates are functions of the modulation $\theta_j=2\pi\beta j+\phi$ (see Eq.~\eqref{eq:hopping_rates}).
Suppose that the rate vanishes at some $\theta = \theta_*$, then $W$ can be expanded to the leading order near $\theta_*$ as $W\propto|\theta-\theta_*|^{2p}$, where $p >0$. As a result for $p\geq1/2$, $\langle1/W\rangle$ diverges and $D_{\mathrm{eff}}$ vanishes.

To characterize transport in this non-diffusive regime, we extract the transport exponent $\nu$ of the mean-square displacement (MSD), $\langle\Delta x^2(t)\rangle\sim t^{\nu}$, from the finite-size scaling of the spectral gap $\Delta_W$~\cite{Mori2024,Znidaric2015}. For a particle spreading across a chain of length $L$, the MSD reaches $ L^2$ asymptotically. 
Since $\tau_{\rm rel}$ is the asymptotic relaxation time of the system, $\langle\Delta x^2(\tau_{\rm rel})\rangle \sim L^2$.
Equating this to $\tau_{\rm rel}^{\nu}$ ($=1/\Delta_W$) yields $\Delta_W\propto L^{-z}$ with dynamical exponent $z=2/\nu$ \cite{prosen2009matrix,Znidaric2015}. This suggests that the transport is subdiffusive for $z>2$, diffusive for $z=2$ and superdiffusive for $z<2$.
 
We relate the spectral gap to the total resistance of the chain, $R_L=\sum_{j=1}^{L-1}W_{j,j+1}^{-1}$.
Excluding rare phases that concentrate resistance towards the end of the chain, which do not affect the exponents as $L\to\infty$, the bounds give $\Delta_W\propto(LR_L)^{-1}$ (End Matter A). The problem therefore reduces to how $R_L$ grows with $L$.

The resistance is dominated by the smallest transfer rates~\cite{Alexander1981,Bouchaud1990Anomalous,Akimoto2020Trace,Monthus2002Localization}, where $\theta_j$ lies closest to the zero $\theta_*$, near which $W^{-1}\propto|\theta_j-\theta_*|^{-2p}$. Since $\beta$ is badly approximable by rationals \cite{kuipers2012uniform,rosenblatt1995pointwise}, $L$ different $\theta_j$ are spread uniformly over the unit circle, with all spacings of order $1/L$~\cite{Slater1967,badziahin2023badly}. Excluding the rare phases that place a bond much closer to $\theta_*$ than $1/L$, the $k$th closest bond therefore lies at a distance of order $k/L$ from $\theta_*$ and contributes a resistance of order $(L/k)^{2p}$. These contributions sum to $R_L\simeq L^{2p}\sum_k k^{-2p}$, which for $p>1/2$ becomes $R_L\propto L^{2p}$ since the series converges. Combined with $\Delta_W\propto(LR_L)^{-1}$, this gives $\Delta_W\propto L^{-(2p+1)}$ and the exponent $\nu=2/(2p+1)$. At the marginal order $p=1/2$, the harmonic series grows logarithmically, $R_L\propto L\ln L$, and transport is diffusive up to logarithmic corrections, $\langle\Delta x^2\rangle\propto t/\ln t$ (End Matter A). In what follows, we apply this general analysis to two specific quasiperiodic models introduced earlier.

 \begin{figure}[t]
\centering
\includegraphics[width=\linewidth]{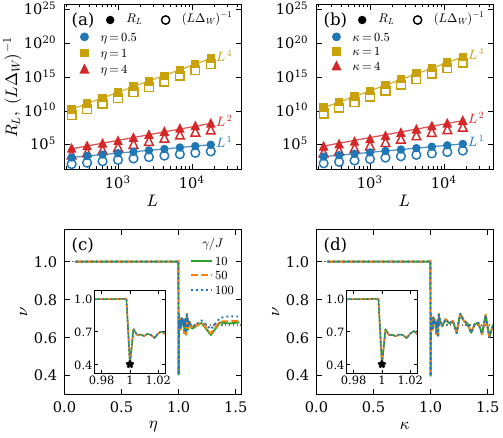}
\caption{Total resistance $R_L$ (filled symbols), evaluated directly using Eq.~\ref{eq:hopping_rates}, and $(L\Delta_W)^{-1}$ (open symbols), obtained independently from numerical diagonalization of $W$ are plotted  versus $L$ for $\gamma=10J$: (a) onsite modulation at $\eta=0.5$, $1$, and $4$ for fixed $\alpha=0.2$ and (b)  hopping modulation at $\kappa=0.5$, $1$, and $4$ in units of $W_0^{-1}$. Note that $W_0=2J^2/\gamma$ and $W_0=2A^2/\gamma$ for (a) and (b), respectively. The transport exponent $\nu=2/z$ obtained from the scaling $\Delta_W\propto L^{-z}$ are plotted as a function of (c) $\eta=\lambda\alpha$ and (d) $\kappa$ for $\gamma/J=10$, $50$, and $100$. Insets in both (c) and (d) are the zoomed in portions at the trench $
\nu=2/5$ (star). }
\label{fig:2}
\end{figure}

\textit{Onsite modulation}--- For the onsite modulation, Eq.~\eqref{eq:Deff_general} gives
\begin{equation}
\frac1{D_{\mathrm{eff}}}=\frac1{D_0}\left[1+\frac{\langle(\Delta V)^2\rangle}{\gamma^2}\right],
\label{eq:Deff_general_analytical}
\end{equation}
where $D_0=2J^2/\gamma$ and $\langle (\Delta V )^2\rangle$ is the average of squared potential difference across nearest-neighbour sites. In the bounded regime ($\eta<1$), the expression can be evaluated analytically. Defining $u=[1-\sqrt{1-\eta^2}]/\eta$ and $\mathcal D(u,\beta)=1-2u^2\cos(2\pi\beta)+u^4$, we obtain the diffusion constant
\begin{equation}
\frac{D_{\mathrm{eff}}}{D_0}=\left[1+\frac{8u^2(1+u^2)\sin^2(\pi\beta)}{(\gamma\alpha)^2(1-\eta^2)(1-u^2)\mathcal D(u,\beta)}\right]^{-1},
\label{eq:Deff_analytical}
\end{equation}
understood continuously at $\eta=0$ (End Matter B). As $\eta\to1^-$,  $D_{\mathrm{eff}}/D_0\simeq\sqrt2\,(\gamma\alpha)^2(1-\eta)^{3/2}$, vanishing continuously at $\eta=1$ for all $\gamma$ which is shown in Fig.~\ref{fig:1}(a). However, a completely different picture emerges for $\eta \geq 1$, where the nature of transport is determined by the order of singularity in potential.

At $\eta=1$, the potential $V(\theta) \propto (1-\cos(\theta))^{-1} \sim \theta^{-2}$ develops a second-order pole at $\theta_* = 0$. Since $V(\theta) \sim \theta^{-2}$ near $\theta_*$, Eq.~\eqref{eq:hopping_rates} gives $W(\theta) \sim \theta^4$ near $\theta_* =0$. By the resistance criterion derived above, the transport exponent is thus $\nu = 2/(2p+1) = 2/5$. Similarly for $\eta >1$, $V(\theta)$ has first-order poles at $\theta_* = \pm \cos^{-1}(1/\eta)$. As a result, $W(\theta) \sim (\theta - \theta_*)^2$ giving $2p=2$ and hence $\nu = 2/(2p+1) = 2/3$. 

To complement this analytical inferences, we numerically obtain the exponents from the scaling up of $\Delta_W$ obtained by diagonalizing $W$. In Fig.~\ref{fig:2}(a), we plot $R_L$ (filled symbols) and $(L\Delta_W)^{-1}$ (open symbols) as a function of $L$ for different values of $\eta$.
The curves follow $R_L\propto L$ and $\Delta_W\propto L^{-2}$ in the diffusive regime and $R_L\propto L^{2p}$ and $\Delta_W\propto L^{-(2p+1)}$ in the subdiffusive regime, agreeeing well with analytical predictions. The slight offset between the two quantities is due to different proportionality constants in two cases. From this analysis we extract the transport exponent $\nu$ for different values of $\eta$ which are shown in Fig.~\ref{fig:2}(c). Plots for different values of $\gamma/J$ show a clear loss of diffusion after $\eta\geq 1$. While for $\eta>1$ the exponent $\nu \sim 2/3$, at $\eta=1$, the exponent becomes $\nu \sim 0.41$ which matches well with the analytically obtained $\nu= 2/5$. Because the exponents are independent of $\gamma$, the trench appears across all dephasing rates [Fig.~\ref{fig:2}(c)]. Repeating this analysis we obtain $\nu$ for different values of $\alpha$ which results in a phase diagram shown in Fig. ~\ref{fig:1}(b). The phase diagram clearly depicts the regions of diffusive (yellow), subdiffusive (green) and the subdiffusive trench (red line) along $\eta=1$ where $\nu\sim 2/5$.

\begin{figure}[t]
\centering
\includegraphics[width=\linewidth]{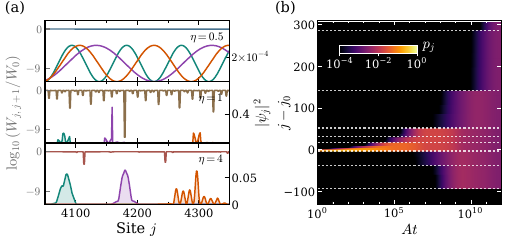}
\caption{(a) Transfer rates $W_{j,j+1}/W_0$ (left axis) and spatial profile $|\psi_j|^2$ of three eigenvectors of the rate matrix $W$ (right axis) against site index $j$ for the onsite model at $\eta=0.5$ (top), $1$ (middle), and $4$ (bottom), with $L=6765$, $\alpha=0.2$, $\gamma=100J$, $\phi=0$, and periodic boundaries.  Eigenvector indices chosen are $n=75$, $35$, $55$ ($\eta=0.5$); $n=1173$, $899$, $1067$ ($\eta=1$); $n=6687$, $6621$, $6229$ ($\eta=4$). Data from sites $4050\le j<4350$ are shown for clarity.(b) Exact Lindblad populations $p_j(t)$ versus $t$ in units of $A^{-1}$ for the hopping model at $\kappa=1$, with $\gamma=10A$ for a system of  size $L=610$ with open boundary condition, and $\phi=5.395$. The initial state  is $\rho(0)=|j_0\rangle\langle j_0|$ with $j_0=L/2$. Dashed lines mark the  bottleneck bonds $2.7\times10^{-13}\le W_{j,j+1}/W_0\le4.1\times10^{-5}$, computed from Eq~\eqref{eq:hopping_rates} .}
\label{fig:3}
\end{figure}

To understand the physical mechanism behind this subdiffusive transport, it is instructive to look at what happens in the real space.
For $\eta\geq1$, isolated potential singularities plunge the rates to near-zero values, generating box-like barriers [Fig.~\ref{fig:3}(a)]. The eigenmodes of $W$ are confined between two such barriers which can be clearly seen from Fig.~\ref{fig:3}(a) where we plot $|\psi_j|^2$ for three exemplary eigenmodes for different values of $\eta$. While for $\eta=0.5$ (top), all the states are delocalized, the states are localized for $\eta=1.0$ (middle) and $\eta=4$ (bottom). Eigenmodes are sharply localized precisely at the trench  ($\eta = 1$) . This slows the spreading of an excitation by confining it between two bottlenecks for long times.

\textit{Hopping modulation}---For the off-diagonal quasiperiodic model,  Eq.~\eqref{eq:hopping_rates} becomes $W_{j,j+1}=2J_j^2/\gamma$. For $\kappa<1$ the hoppings are bounded away from zero, and integrating the bond resistance gives the exact diffusion constant (End Matter B)
\begin{equation}
\frac{D_{\mathrm{eff}}}{D_0}=(1-\kappa^2)^{3/2},\qquad D_0=\frac{2A^2}{\gamma},
\label{eq:Deff_offdiag}
\end{equation}
which vanishes as $(1-\kappa)^{3/2}$ on approaching the threshold, with the same exponent $3/2$ as in the onsite model. Expanding the hopping amplitude near $\theta_j=\pi$ gives
\begin{equation}
\frac{J_j}{A}=1-\kappa+\frac\kappa2(\theta_j-\pi)^2+O[(\theta_j-\pi)^4].
\label{eq:hopping_merger}
\end{equation}
At $\kappa=1$ the hopping has a quadratic zero and $W\sim(\theta_j-\pi)^4$ leading to $p=2$ and hence the subdiffusive trench $\nu=2/5$. At any fixed $\kappa>1$ it has two simple zeros and $W$ vanishes quadratically, giving $p=1$ and $\nu=2/3$. These findings are  confirmed numerically in Fig.~\ref{fig:2}(d).

Since the hopping amplitude vanishes for $\kappa\geq1$, bonds with $J_j=0$ would be expected to cut the chain into disconnected pieces and trivially localize the particle~\cite{Longhi2024}. However, our analysis shows that for irrational $\beta$, a bond vanishes exactly only for a measure-zero set of phases $\phi$, and even then at most one bond per zero, so the chain is generically never broken into finite segments. Instead, as the chain grows, its weakest bonds approach zero arbitrarily closely without vanishing. These near-zero bonds act as the bottlenecks of box-confined relaxation, which is the origin of the subdiffusive transport.

To illustrate this box confined transport, we show the spreading of an excitation in the lattice in Fig.~\ref{fig:3}(b). The excitation spreads diffusively inside a box, encounters a barrier, and remains trapped for long times before escaping to the adjacent box, thereby slowing down the transport (for more details, see End Matter D). Similar box-confined transport is also seen 
for onsite modulation (not shown).
 
\begin{figure}[t]
\centering
\includegraphics[width=\linewidth]{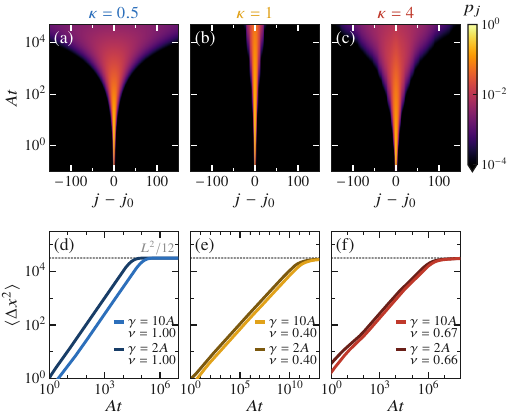}
\caption{Exact Lindblad dynamics of the hopping model on an open chain of $L=610$ sites, with the particle initially at the central site and averaged over $600$ phases $\phi$. (a)-(c) Populations $p_j(t)$ at $\gamma=10A$: the packet spreads diffusively at $\kappa=0.5$, and subdiffusively at $\kappa=1$ and $4$. (d)-(f) Mean-square displacement $\langle\Delta x^2\rangle$ for $\gamma=10A$ (light) and $\gamma=2A$ (dark), up to saturation at $L^2/12$ (dotted line). Power-law fits give $\nu=1.00$, $0.40$, and $0.67$ at $\gamma=10A$ and $\nu=1.00$, $0.40$, and $0.66$ at $\gamma=2A$, in agreement with the predicted values $1$, $2/5$, and $2/3$.}
\label{fig:4}
\end{figure}

\textit{Exact Lindblad dynamics}---To rule out artifacts of the quantum-to-classical mapping, we solve the full Lindblad equation~\eqref{eq:LME} for the hopping model. For a single particle, uniform dephasing acts as a sequence of position measurements at exponentially distributed times~\cite{HofmannDrossel2021,LozanoNegro2024}, so the site populations obey an exact renewal equation, valid for any $\gamma/J$. We solve this equation in Laplace space and invert the transform numerically, which yields the exact populations without time-stepping (End Matter C). For a particle initially at the center of a chain of $L=610$ sites, averaged over 600 phases, the density spreads diffusively at $\kappa=0.5$, remains tightly confined at $\kappa=1$, and spreads subdiffusively at $\kappa=4$ [Fig.~\ref{fig:4}(a)-(c)]. The MSD follows a power law growth in time before saturation, with $\nu=1.00$, $0.40$, and $0.67$ [Fig.~\ref{fig:4}(d)-(f)], in agreement with the predicted $1$, $2/5$, and $2/3$; increasing $L$ only extends the power-law regime (End Matter D).

Reducing the dephasing rate to $\gamma=2A$, well outside the strong-dephasing regime, where some bonds of the $\kappa=4$ chain exceed $\gamma$, shifts the curves in time but leaves the exponents unchanged, $\nu=1.00$, $0.40$, and $0.66$ [Fig.~\ref{fig:4}(d)-(f)]. This robustness follows from the mechanism: the transport is limited by the weakest bonds, where $|J_j|\ll\gamma$ holds for any finite dephasing rate.
 
\textit{Discussion}| The transport phenomena discussed here extend beyond the specific models considered. If a potential has singularity of order $2m$, the resulting rate $W$ has a zero of order $2p=4m$ [Eq.~\ref{eq:hopping_rates}], predicting an exponent $\nu=2/(2p+1) = 2/(4m+1)$. Likewise, a hopping zero of order $q$ gives $p=q$ and $\nu=2/(2q+1)$. For example, potentials with simple poles, such as the Maryland model $V_j\propto\tan(\pi\beta j+\phi)$~\cite{Grempel1982}, have $p=1$ and are therefore subdiffusive with $\nu=2/3$ for any strength of the potential. Conversely, neither an unbounded potential nor a vanishing hopping by itself destroys diffusion. A singularity of the potential, $V\sim|\theta-\theta_*|^{-s}$, or a zero of the hopping, $J\sim|\theta-\theta_*|^{s}$, of order $s<1/2$ leaves $\langle 1/W \rangle$ finite and the transport remains diffusive.

Modulating the hopping offers a promising experimental platform to observe the transport transition. In photonic quantum walks in synthetic mesh
lattices with stochastic phase modulation \cite{Pankov:19,Longhi2024,Regensburger2011Photon,Cedzich2024Mobility,Broome2010Discrete,Schreiber2011Decoherence}, the incoherent transfer probability across bond $n$ is
$\cos^2\beta_n \propto J_n^2$~\cite{Longhi2024} for coupling angles $\beta_n$ near $\pi/2$, so tuning $\kappa$ through $1$
realizes the trench, and the exponents $\nu = 2/5$ and $2/3$ can be read off from the spreading of
light pulses. 
In principle, higher-order trenches may also arise for more singular hopping profiles; for instance, quartic zeros of the hopping amplitudes, as realized by $J_j=[1+\kappa\cos(2\pi\beta j+\phi)]^2$, give rise to a trench with $\nu=2/9$.
Synthetic momentum lattices of ultracold atoms \cite{Xiao2021, An2017Ballistic}, in which the coherent model has been
realized, offer a complementary platform once controlled noise is added. 
 
\textit{Conclusion}---We have addressed a fundamental question: when does dephasing lead to subdiffusion in a quasiperiodic lattice? 
Mapping the dephased dynamics onto a classical random walk, we showed that diffusion survives if and only if the mean resistance of the chain remains finite, a condition broken by singularities of the onsite potential or zeros of the hopping. Diffusion coefficient vanishes continuously as we approach the threshold modulation; at and beyond it, the transfer rates become arbitrarily small but generically never vanish. The weakest bonds form bottlenecks and confine relaxation to boxes. The resulting subdiffusion has $\nu=2/(2p+1)$, set solely by the order $2p$ of the rate zeros. Since the zero at the threshold is of higher order, transport is slowest precisely at the onset of subdiffusive transport, forming a subdiffusive trench that is independent of the dephasing rate and persists in exact Lindblad simulations even at dephasing rate $\gamma=2A$, which is well outside the strong-dephasing regime.

This finding resembles bottleneck-induced subdiffusion, known in classical transport, where the exponent is set by an imposed distribution of independent random rates~\cite{Alexander1981,Bouchaud1990Anomalous,HausKehr1987}. 
However, in our case, similar behavior emerges in quantum lattices with deterministic modulation: no randomness is put in by hand, the rates are correlated, and the bottlenecks are generated by the Hamiltonian itself. The exponents are then set by the modulation and tunable through a single parameter, producing a non-monotonic subdiffusive trench with no natural counterpart in classical models. Dephased quasiperiodic lattices thus provide a deterministic setting in which bottlenecks can be engineered rather than drawn at random, which are of interest to quantum and classical transport communities. Whether this criterion holds in higher dimensions, where alternative paths may bypass a bottleneck~\cite{Kirkpatrick1973,HavlinBenAvraham2002}, or in interacting systems, where collective pathways may arise~\cite{Gopalakrishnan2017,Medvedyeva2016,TaylorScardicchio2021}, remains an open question.

\paragraph{Acknowledgement.-} T.M. acknowledges support from Anusandhan National Research Foundation (ANRF), Government of India, through project No. ANRF/ARG/2025/007442/PS..

\paragraph{Data availability.-} The data and code supporting the findings of this work are available from the authors upon reasonable request.

\bibliographystyle{apsrev4-2}
\bibliography{references.bib}

\clearpage
\twocolumngrid
\section*{End Matter}
\textit{A. Variational Bounds on Spectral Gap}---Consider a chain of $L$ sites with bond rates $w_j\equiv W_{j,j+1}>0$. Treating each bond as a resistor $w_j^{-1}$, the total resistance is $R_L=\sum_{j=1}^{L-1}w_j^{-1}$, and $f_j=R_L^{-1}\sum_{k<j}w_k^{-1}$ is the fraction of this resistance accumulated between site $1$ and site $j$.

Since \(W\) is symmetric and has a uniform zero mode $\mathbf{1}$, the variational principle applied to $-W$ in the subspace orthogonal to $\mathbf{1}$ gives~\cite{horn2012matrix,sakurai1994modern}
\begin{equation}
\Delta_W = |\lambda_1|=\min_{x\perp\mathbf{1}}\frac{\langle x|-W|x\rangle}{\langle x|x\rangle}.
\end{equation}
Using \(\langle x|-W|x\rangle=\sum_j w_j(x_{j+1}-x_j)^2\), and expressing the orthogonality constraint through subtraction of the mean, we obtain
\begin{equation}
\Delta_W=\min_{x\not\equiv\mathrm{constant}}
\frac{\sum_j w_j(x_{j+1}-x_j)^2}{\sum_i(x_i-\bar x)^2}. \label{eq: Rayleigh-Ritz Bound}
\end{equation}
Using Cauchy-Schwarz inequality,
\begin{align}
    (x_k-x_i)^2& =\left(\sum_{j=i}^{k-1}w_j^{-1/2}\,w_j^{1/2}(x_{j+1}-x_j) \right)^2 \nonumber \\ & \leq R_L\sum_j w_j(x_{j+1}-x_j)^2
\end{align}
Summing over $L(L-1)/2$ unique $i,k$ combinations on both sides and using $\sum_{i<k}(x_k-x_i)^2=L\sum_i(x_i-\bar x)^2$ gives the bound
\begin{equation}
    L \sum_i (x_i - \bar{x})^2 \leq \frac{L(L-1)}{2} R_L \sum_j w_j (x_{j+1} - x_{j})^2
\end{equation}
Combined with Eq.~\eqref{eq: Rayleigh-Ritz Bound} gives the lower bound
\begin{equation}
\frac{2}{(L-1)R_L}\leq\Delta_W\leq\frac{1}{R_L\sum_{j}(f_j-\bar f)^2},
\label{eq:gap_bounds}
\end{equation}
and the upper bound follows from choosing $x_j = f_j$ giving $\sum_j w_j (f_{j+1} -f_j)^2 = \sum_j w_j w_j^{-2}/R_L^2 = 1/R_L$.

The sum in the upper bound is the spatial variance of the resistance profile $f_j$. A resistance profile dominated by bottlenecks in the bulk produces variations of $f_j$ across an extensive fraction of the chain, giving $\sum_j(f_j-\bar f)^2=O(L)$. The phases placing the bottlenecks near the boundaries form a vanishing fraction in the thermodynamic limit, and are hence ignored. Consequently, both bounds scale as $(LR_L)^{-1}$, giving $\Delta_W \propto (LR_L)^{-1}$. Combined with $R_L \propto L^{2p}$, this gives $\Delta_W \propto L^{-2p-1}$ for $p>1/2$. At the marginal order $p=1/2$, the harmonic series diverges, producing logarithmic corrections to diffusion, giving $\langle\Delta x^2\rangle\propto t/\ln t$ \cite{bertin2003subdiffusion}.

\textit{B. Diffusion coefficients}---By Eq.~\eqref{eq:Deff_general}, $D_{\mathrm{eff}}=\langle1/W\rangle^{-1}$. Since the phases $\theta_j$ are equidistributed, the chain average of any integrable function of $\theta_j$ equals its phase average in the thermodynamic limit, $\langle g\rangle=\int_0^{2\pi}\frac{d\theta}{2\pi}\,g(\theta)$~\cite{kuipers2012uniform,weyl1916gleichverteilung}.

\emph{Hopping modulation.} With $V_j=0$ and $J_j=A(1+\kappa\cos\theta_j)$, $1/W_{j,j+1}=\gamma/[2A^2(1+\kappa\cos\theta_j)^2]$. Differentiating $\int_0^{2\pi}\frac{d\theta}{2\pi}(a+b\cos\theta)^{-1}=(a^2-b^2)^{-1/2}$ \cite{gradshteyn2014table} with respect to $a$ and setting $a=1$, $b=\kappa$ gives, for $\kappa<1$,
\begin{equation}
\Bigl\langle\frac1W\Bigr\rangle=\frac{\gamma}{2A^2}\int_0^{2\pi}\frac{d\theta}{2\pi}\frac{1}{(1+\kappa\cos\theta)^2}=\frac{1}{D_0(1-\kappa^2)^{3/2}},
\end{equation}
with $D_0=2A^2/\gamma$, which is Eq.~\eqref{eq:Deff_offdiag}. As $\kappa\to1^-$, $D_{\mathrm{eff}}/D_0\simeq2^{3/2}(1-\kappa)^{3/2}$, vanishing at $\kappa = 1$.

\emph{Onsite modulation.} With $J_j=J$, Eq.~\eqref{eq:hopping_rates} gives $1/W_{j,j+1}=[1+(\Delta V_j)^2/\gamma^2]/D_0$ with $D_0=2J^2/\gamma$, which yields Eq.~\eqref{eq:Deff_general_analytical}. Writing $\eta=\lambda\alpha$, the potential is $V_j=(F_j-1)/\alpha$ with $F(\theta)=(1-\eta\cos\theta)^{-1}$. For $\eta<1$, define $u=(1-\sqrt{1-\eta^2})/\eta\in[0,1)$. Then $1-\eta\cos\theta=(1-2u\cos\theta+u^2)/(1+u^2)$, and the Poisson-kernel expansion \cite{stein2011fourier} gives 
\begin{equation}
F(\theta)=\frac{1}{\sqrt{1-\eta^2}}\Bigl[1+2\sum_{n\geq1}u^n\cos n\theta\Bigr].
\end{equation}
Since $\theta_{j+1}=\theta_j+2\pi\beta$,
\begin{equation}
\Delta V_j=V_j-V_{j+1}=\frac{4}{\alpha\sqrt{1-\eta^2}}\sum_{n\geq1}u^n\sin(n\pi\beta)\sin[n(\theta_j+\pi\beta)].
\end{equation}
Averaging the square over $\theta$, only diagonal terms survive, and summing the series \cite{gradshteyn2014table},
\begin{align}
\langle(\Delta V)^2\rangle& =\frac{8}{\alpha^2(1-\eta^2)}\sum_{n\geq1}u^{2n}\sin^2(n\pi\beta) \\&
=\frac{8u^2(1+u^2)\sin^2(\pi\beta)}{\alpha^2(1-\eta^2)(1-u^2)\,\mathcal D(u,\beta)},
\end{align}
with $\mathcal D(u,\beta)=1-2u^2\cos(2\pi\beta)+u^4$, which gives Eq.~\eqref{eq:Deff_analytical}. For $\eta\to0$, $u\simeq\eta/2$ and $\langle(\Delta V)^2\rangle\to2\lambda^2\sin^2(\pi\beta)$, the Aubry-Andr\'e result. For $\eta\to1^-$, $u\simeq1-\sqrt{2(1-\eta)}$ and $\mathcal D\to4\sin^2(\pi\beta)$, so $\langle(\Delta V)^2\rangle\simeq(\sqrt2\alpha^2)^{-1}(1-\eta)^{-3/2}$ and $D_{\mathrm{eff}}/D_0\simeq\sqrt2(\gamma\alpha)^2(1-\eta)^{3/2}$. For $\eta\geq1$, $(\Delta V)^2$ is not integrable near the poles and $D_{\mathrm{eff}}=0$.

\begin{figure}[t]
    \centering
    \includegraphics[width=\linewidth]{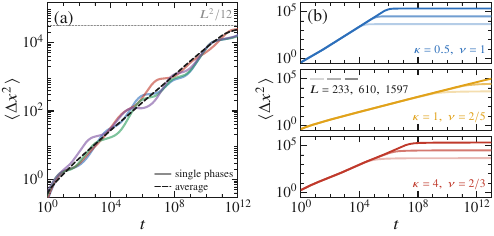}
    \caption{Exact Lindblad dynamics of the hopping model at $\gamma=10A$, with the particle initially at the central site. (a) $\langle\Delta x^2\rangle$ at $\kappa=1$ and $L=610$ for four individual phases $\phi$ (colored) and the average over 600 phases (dashed). For a single phase, plateaus in $\langle\Delta x^2\rangle$ reflect box-confined relaxation. Averaging over $\phi$ smooths the staircases into $\langle\Delta x^2\rangle\propto t^{2/5}$ up to saturation. (b) Phase-averaged $\langle\Delta x^2\rangle$ for $\kappa=0.5$, $1$, and $4$ (top to bottom) and $L=233$, $610$, and $1597$ (light to dark; 600, 600, and 300 phases). All sizes follow the same power law $t^{\nu}$, with $\nu=1$, $2/5$, and $2/3$, until saturation, showing that the exponents are asymptotic.}
    \label{fig:5}
\end{figure}

\textit{C. Exact Lindblad dynamics.}---For a single particle, $\sum_j\hat n_j\hat\rho\hat n_j\equiv\Phi[\hat\rho]$ is the diagonal part of $\hat\rho$, so Eq.~\eqref{eq:LME} takes the form $d\hat\rho/dt=(\mathcal L_0+\gamma\Phi)\hat\rho$ with $\mathcal L_0\hat\rho=-i[\hat H,\hat\rho]-\gamma\hat\rho$~\cite{HofmannDrossel2021}. Treating $\gamma\Phi[\hat\rho]$ as a source term, Duhamel's formula~\cite{Tao2006} gives the exact relation
\begin{equation}
\hat\rho(t)=e^{\mathcal L_0t}\hat\rho_0+\gamma\int_0^t\!ds\,e^{\mathcal L_0s}\,\Phi[\hat\rho(t-s)].
\label{eq:duhamel}
\end{equation}
We apply Eq.~\eqref{eq:duhamel} to $\hat\rho_0=|j_0\rangle\langle j_0|$ and take the diagonal. Since $\Phi[\hat\rho(t-s)]$ contains only the populations $p(t-s)$, and $e^{\mathcal L_0s}$ maps $\hat\rho\mapsto e^{-\gamma s}\hat U(s)\,\hat\rho\,\hat U^\dagger(s)$ with $\hat U(s)=e^{-i\hat Hs}$, the populations obey the closed renewal equation
\begin{equation}
p(t)=e^{-\gamma t}M(t)\,p_0+\gamma\!\int_0^t\!ds\,e^{-\gamma s}M(s)\,p(t-s),
\label{eq:renewal}
\end{equation}
where $M_{kj}(s)=|\langle k|\hat U(s)|j\rangle|^2$. This is the population-level form of coherent evolution interrupted by dephasing events at exponentially distributed times~\cite{HofmannDrossel2021,Vacchini2020}.

Under the Laplace transform $\hat p(z)=\int_0^\infty dt\,e^{-zt}p(t)$, the convolution becomes a product, and Eq.~\eqref{eq:renewal} gives
\begin{equation}
\hat p(z)=\frac{1}{\gamma}\left([1-A(z)]^{-1}-1\right)p_0,\qquad A(z)=\gamma\hat M(z+\gamma).
\label{eq:laplace}
\end{equation}
Inserting the spectral decomposition $\hat H=\sum_mE_m|\psi_m\rangle\langle\psi_m|$, with real eigenvectors $V_{km}=\langle k|\psi_m\rangle$, gives
\begin{equation}
M_{kj}(s)=\sum_{mn}V_{km}V_{kn}V_{jm}V_{jn}\,e^{-i(E_m-E_n)s},
\end{equation}
whose Laplace transform at $z+\gamma$ yields
\begin{equation}
A_{kj}(z)=\gamma\sum_{mn}V_{km}V_{kn}V_{jm}V_{jn}\,\frac{z+\gamma}{(z+\gamma)^2+(E_m-E_n)^2}.
\end{equation}
We evaluate Eq.~\eqref{eq:laplace} exactly at complex $z$ and invert the Laplace transform numerically~\cite{Weideman2007,Trefethen2014,LopezFernandez2006}, which yields $p(t)$ without time-stepping at arbitrarily late times. The results agree with exact diagonalization of the full Liouvillian for $L=34$ to $4\times10^{-9}$ over $10^{-2}\leq At\leq10^{12}$.

\textit{D. Box-confined relaxation and finite-size scaling}---Figure~\ref{fig:5} resolves the exact Lindblad dynamics of the hopping model behind Fig.~\ref{fig:4}. For individual phases at $\kappa=1$ [Fig.~\ref{fig:5}(a)], $\langle\Delta x^2\rangle$ grows as a staircase: on each plateau the packet fills a box bounded by nearly vanishing bonds, and each step marks its escape across the next bottleneck. Averaging over $\phi$ smooths them into the power law $t^{2/5}$. Figure~\ref{fig:5}(b) shows the phase-averaged MSD for $L=233$, $610$, and $1597$: all sizes follow the same power law until saturation, and increasing $L$ only extends its range, so the exponents $\nu=1$, $2/5$, and $2/3$ are asymptotic rather than finite-size transients.

\end{document}